\documentclass[psfig]{mn2e}

\usepackage{times}

\newcommand{\bgr}{\bibitem[\protect\citename{dummy }1893]{dum}}
\newcommand{\etalc}{et~al.}
\newcommand\xmm{{\em XMM-Newton\/ }}

\begin{document}
\title[The nature of X-ray selected EROs]
{The nature of X-ray selected EROs}
{}

\author[J. A. Stevens et al.]
{J.\ A.\ Stevens$^{1,2}$, M.\ J.\ Page$^2$, R.\ J.\ Ivison$^1$, 
Ian\ Smail$^3$, I.\ Lehmann$^{4,5}$, G.\ Hasinger$^{4,5}$ \newauthor and G.\
Szokoly$^{4,5}$ 
\\
$^1$ Astronomy Technology Centre, Royal Observatory, Blackford Hill,
Edinburgh EH9 3HJ \\
$^2$ Mullard Space Science Laboratory, University College London, Holmbury
St. Mary, Dorking, Surrey, RH5 6NT  \\
$^3$ Department of Physics, University of Durham, South Road, Durham 
DH1 3LE \\ 
$^4$ Astrophysikalisches Institut Potsdam (AIP), An der Sternwarte 16, 14482
Potsdam, Germany \\
$^5$ Max-Plank-Institut F\"{u}r extraterrestrische Physik, Giessenbachstrasse
PF 1312, 85748 Garching bei Muenchen, Germany \\
}
\date{draft 1.0}

\maketitle
\begin{abstract}
\noindent
We report on the X-ray, optical, near-infrared, submillimetre and radio
properties of five Extremely Red Objects (EROs) selected at X-ray wavelengths
by {\em XMM-Newton\/} in the Lockman Hole field. They all have enough counts in
the X-ray band to allow spectral fitting: four are most probably obscured,
Compton-thin AGN with redshift dependent absorbing column densities of
$10^{22}-10^{24}$~cm$^{-2}$, whilst the fifth is best fitted by a thermal
spectrum and is likely to be a massive elliptical galaxy in a deep
gravitational potential.  Their optical/near-infrared colours and sizes suggest
that X-ray selected EROs comprise a mixture of dusty `starburst' galaxies and
non-dusty galaxies that are dominated by either star-light or light from an
active nucleus. The colour diagnostics are supported by the submillimetre and
radio data: the two AGN with `starburst' colours have submillimetre or radio
flux densities that imply large star-formation rates, whilst those with
`elliptical' colours do not.  The one source detected in the submillimetre
waveband has narrow emission lines at a redshift of 1.45.  Although the
bulk of its radio emission originates from processes other than star formation,
it is most probably a radio-quiet ultraluminous infrared galaxy.

\end{abstract}

%

\section{Introduction}

\begin{table*}
\footnotesize
\centering
\def\baselinestretch{1}                                     
\caption[dum]{\small Source names and coordinates of optical
counterparts. Sources with $Rosat$ numbers were detected in the previous deep
survey of the Lockman Hole. Source number 84 is discussed by Lehmann et
al. (2001).}
\label{table:names}
\vspace*{0.1in}
\begin{tabular}{ccccc}
ero\# & Source name & RA & Dec & {\em Rosat\/} number \\ 
      &             & J2000 & J2000 &                      \\ \\
ero1 & XMMUJ $105216.9+572017$ & $10\ 52\ 16.97$ & $+57\ 20\ 17.1$ & $84$ \\
ero2 & XMMUJ $105348.1+572817$ & $10\ 53\ 48.05$ & $+57\ 28\ 16.7$ & \ldots \\
ero3 & XMMUJ $105146.5+573037$ & $10\ 51\ 46.58$ & $+57\ 30\ 35.1$ & \ldots \\
ero4 & XMMUJ $105225.5+572551$ & $10\ 52\ 25.29$ & $+57\ 25\ 50.7$ & $491$ \\
ero5 & XMMUJ $105255.5+571950$ & $10\ 52\ 55.50$ & $+57\ 19\ 51.0$ & $438$ \\
\multicolumn{3}{l}{}
\end {tabular}
\end{table*}

\begin{table*}
\footnotesize
\centering
\def\baselinestretch{1}                                     
\caption[dum]{\small X-ray fluxes in the various bands. Column 2 has the
source circle, $\o$, used to extract the X-ray spectrum. Column 3 has the total
source counts within this circle.}
\label{table:posx}
\vspace*{0.1in}
\begin{tabular}{ccccccc}
Name & $\o$ & counts & $S_{0.2-0.5\,{\rm keV}}$ & $S_{0.5-2.0\,{\rm
keV}}$ & $S_{2.0-5.0\,{\rm keV}}$ &  $S_{5.0-12\,{\rm keV}}$ \\ 
& arcsec & [$0.2-12$ keV] & \multicolumn{4}{c}{$10^{-16}$erg\ cm$^{-2}$\ s$^{-1}$} \\ \\
ero1 & $30.0$ & $504$ &$0.64\pm5.32$ & $56.68\pm4.07$ & $170.42\pm11.45$ & $378.18\pm70.41$ \\ 
ero2 & $19.0$ & $153$ &$0.00\pm4.48$ & $24.97\pm3.24$ & $53.30\pm7.71$ & $224.09\pm62.78$ \\
ero3 & $14.5$ & $132$ &$17.27\pm3.26$ & $32.30\pm3.34$ & $19.51\pm5.01$ & $61.67\pm45.37$ \\
ero4 & $13.0$ & $244$ &$1.68\pm1.21$ & $22.11\pm2.27$ & $60.38\pm5.77$ & $180.43\pm37.51$ \\
ero5 & $22.4$ & 88  &$1.53\pm1.40$ & $17.85\pm2.90$ & $22.17\pm5.75$ & $116.87\pm57.01$ \\
\multicolumn{3}{l}{}
\end {tabular}
\end{table*}

The recent launch of the \xmm and {\em Chandra\/} observatories has produced
considerable interest in the nature of the faint extragalactic X-ray
population. In particular, the X-ray background at $2 - 10$~keV energies  has been
almost completely resolved for the first time by these facilities. Synthesis
models for the hard X-ray background, which require a large population of
heavily obscured AGN (e.g. Setti \& Woltjer 1989),
predict that about 85 per cent of the accretion power in the Universe is
absorbed (Fabian \& Iwasawa 1999). This absorbed energy is expected to be
re-emitted in the infrared--millimetre waveband.  


One of the most interesting discoveries from the previous, {\em Rosat\/}, era
was that a small number of faint X-ray sources were identified with extremely
red optical/infrared counterparts (Newsam et al. 1997; Lehmann et al. 2001).
Extremely red objects (EROs) were discovered more than a decade ago (Elston,
Rieke \& Rieke 1988) but their connection with the broad extragalactic
population is only just becoming clear. It is thought that the population is
bimodal; EROs are either (1) elliptical galaxies at high-redshift ($z \sim 1$),
their red colours being produced by an evolved stellar population, or (2) they
are starburst galaxies at high-redshift ($z > 1$) and their red colours are
produced by the suppression of UV light from young stars by dust. However,
there is growing evidence that the population of EROs might be more diverse
than originally thought with several sources showing evidence of AGN activity
(Smith et al. 2001; Pierre et al. 2001; Afonso et al. 2001).  In a recent
study, the X-ray colours of EROs in the {\em Chandra\/} Deep Field North
suggest that 9 out of 13 of them have X-ray emission from obscured AGN
(Alexander et al. 2002). The other $4$ are likely to be a mixture of
low-luminosity AGNs, normal ellipticals and starbursts.

The first deep survey made with \xmm was a 100~ks observation of the Lockman
Hole (Hasinger et al. 2001). Follow-up observations revealed a population of
EROs. Most of these EROs have high X-ray-to-optical flux ratios indicating the
presence of an obscured AGN (Mainieri et al. 2002). The large collecting area
of \xmm makes it possible to examine their X-ray spectra on a case by case
basis, providing an excellent opportunity to determine the origin of their
X-ray emission, and thereby investigate the role of buried AGN in these
objects.  To elucidate the connection between the X-ray and global properties
of EROs we here report on radio, millimetre, submillimetre and
optical/near-infrared observations of five objects selected from the 100~ks
\xmm observation of the Lockman Hole.

Throughout this paper we define spectral index, $\alpha$, as
$S_{\nu}\propto\nu^{\alpha}$ where $S_{\nu}$ is the flux density at frequency
$\nu$. We assume $H_0=50$~km\,s$^{-1}$\,Mpc$^{-1}$, and $q_0=0.5$ throughout
unless otherwise stated.


\section{Sample}

\label{sec-sample}

Our sample comprises five out of six of the reddest sources (those satisfying
$R-K'>6$) in the 100~ks \xmm observation of the Lockman Hole (Hasinger et
al. 2001). Due to telescope time constraints, the other source was not observed
in the submillimetre band and, likewise, lacks good quality imaging data in the
near-infrared. For clarity it is not discussed in this paper.  In
Table~\ref{table:names} we give the names and coordinates of the sources; three
of them were detected in the {\em Rosat\/} Deep Survey although they were not
spectroscopically identified (Hasinger et al. 1998; Lehmann et al. 2001).
However, ero5 now has a measured redshift of $z=1.45$ from the detection of
narrow H$\alpha$ in a $K$-band spectrum and narrow [O\,{\sc\small
III}]$\lambda5007$ in a $H$-band spectrum, both taken at the Subaru telescope
(PIs: Masayuki Akiyama, Patrick Henry). In addition, ero1, the hardest source
in the {\em Rosat\/} sample has a photometric redshift of $z\sim2.7$, although
redshifts as low as $1.5$ are statistically acceptable (Lehmann et
al. 2001). However, photometric redshifts should be treated with caution given
that an AGN might contribute to the optical/near-infrared light (see
Sections~\ref{sec-rnir} and \ref{sec-disc}).

In the part of the \xmm Lockman field that has been imaged in the $K$ band
there are 151 objects that satisfy $R-K'>6$ and $K\leq20$. Of these 6 (4 per
cent) have been detected as X-ray sources.


\section{Observations}

\subsection{X-ray observations}

\begin{table*}
\footnotesize
\centering
\def\baselinestretch{1}                                     
\caption[dum]{\small Optical and near-infrared photometry, colours and half-light
radii ($r_{\rm hl}$)}
\label{table:oir}
\vspace*{0.1in}
\begin{tabular}{lcccccccc}
Name & $R$ & $J$ & $t_{\rm int}$ ($J$) & $K$ & $t_{\rm int}$ ($K$) & $R-K$ &
 $J-K$ & $r_{\rm hl}$ \\ 
 & mag & mag & sec & mag & sec & mag & mag & arcsec \\ \\
ero1$^a$ & $25.5$ & $21.85\pm0.12$ & 9720 & $19.45\pm0.04$ & 3240& $6.1\pm0.05$ &
 $2.40\pm0.13$ & $0.12\pm0.08$ \\  
ero2 & $25.5$ & $21.38\pm0.08$ & 6480 & $18.91\pm0.03$ & 2160 & $6.6\pm0.05$ &
 $2.47\pm0.07$ & $0.16\pm0.06$ \\ 
ero3 & $25.6$ & $21.75\pm0.10$ & 12960 & $19.80\pm0.05$ & 4860 & $5.8\pm0.06$ &
 $1.95\pm0.11$ & $0.43\pm0.07$\\
ero4 & $24.6$ & $20.44\pm0.05$ & 9180 & $18.79\pm0.03$ & 2160 & $5.8\pm0.05$ &
 $1.65\pm0.06$ & $0.29\pm0.05$ \\
ero5 & $24.2$ & $20.90\pm0.15$ & 2160 & $18.29\pm0.05$ & 1620 & $5.9\pm0.06$ &
 $2.61\pm0.16$ & $0.00\pm0.08$ \\ \\
\multicolumn{9}{l}{\small $^a$Lehmann et al. (2001) find $K=19.4\pm0.1$ and
 $J=21.7\pm0.2$ for ero1, consistent with the values presented here.}
\end {tabular}
\end{table*}

\begin{table*}
\footnotesize
\centering
\def\baselinestretch{1}                                     
\caption[dum]{\small Radio positions measured from the 1.4-GHz VLA map,  
radio and submillimetre flux measurements. The values quoted in columns 4 and 5
are offsets between the radio positions and those quoted in
Table~\ref{table:posx}. Column 6 has the source size measured at 1.4~GHz; the measurements
have not been deconvolved with the 1.4 arcsec beam.}
\label{table:radsub}
\vspace*{0.1in}
\begin{tabular}{cccccccccc}
Name & RA  & Dec & $|\Delta$RA$|$ & $|\Delta$Dec$|$ & $\theta_{FWHM}$ &$1.4$ GHz & $4.9$ GHz& $450$
     $\mu$m & $850$ $\mu$m \\ 
     &  J2000 & J2000 & arcsec & arcsec & arcsec & $\mu$Jy & $\mu$Jy&  mJy      &  mJy
     \\ \\ 
ero1 & $10\ 52\ 16.90$ & $57\ 20\ 17.2$ & $0.6$ & $0.1$ & $1.4\pm0.1$ &
     $113\pm12$ & $104\pm39$ & $2.2\pm10.5$ & $-1.4\pm1.4$ \\
ero2 & \ldots & \ldots & \ldots & \ldots & \ldots & $3\sigma < 27$ & \ldots&
     $-3.6\pm8.4$ & $1.8\pm1.0$ \\  
ero3 & \ldots& \ldots & \ldots & \ldots & \dots & $3\sigma < 16$ & $3\sigma < 43$&
     $10.6\pm7.5$ & $1.8\pm1.1$ \\ 
ero4 & $10\ 52\ 25.44$ & $57\ 25\ 51.1$ & $1.2$ & $0.4$ & $1.6\pm0.4$ &$31\pm13$
     &  $3\sigma < 35$ & $-9.2\pm22.4$ & $-2.5\pm1.6$ \\ 
ero5 & $10\ 52\ 55.32$ & $57\ 19\ 50.4$ & $1.5$ & $0.6$ & $1.43\pm0.01$
     &$2990\pm100$ & $1090\pm70$& $16.7\pm9.8$ & $4.3\pm1.0$ \\ 
\multicolumn{3}{l}{}
\end {tabular}
\end{table*}

The Lockman Hole was observed five times by the {\em XMM-Newton\/} EPIC cameras
during April and May 2000, as described in Hasinger et al. (2001). The data
presented here are the same as those used by Hasinger et al. (2001) but we have
reprocessed the data with the {\em XMM-Newton\/} Science Analysis System (SAS)
version 5.3, using the latest calibration data. The events were transformed to
a common astrometric frame by cross-correlating the source positions derived
from each of the observations with the optical counterpart positions from the
previous {\em Rosat\/} HRI survey (Lehmann et al. 2001).

Images and exposure maps were formed for each observation in four energy bands:
$0.2-0.5$ keV, $0.5-2.0$ keV, $2.0-5.0$ keV and $5.0-12.0$ keV; photons with
energies close to those of the strong instrumental emission lines (Lumb et
al. 2002) were excluded from the images.  Background maps for each observation
and each energy band were created by maximum-likelihood fitting of uniform
vignetted and unvignetted components to the images after sources had been
excised. Images, exposure maps and background maps from the different
observations were then combined. The images in the four energy bands were
simultaneously searched for sources using the SAS tasks {\small EBOXDETECT} and
{\small EMLDETECT}. To optimize the background determination we performed
several iterations of source detection and background fitting. Count rates were
converted to fluxes assuming a power law spectrum with $\alpha = -0.7$.

For spectral analysis we obtained source counts from circles of between 13
arcsec and 30 arcsec radius centred on the source positions given in
Table~\ref{table:posx}. The sizes of the source circles were determined
primarily by the need to avoid contamination by other nearby
sources. Background counts were obtained from annular regions of diameter 2
arcmin around the source positions; sources with more than 40 counts were
excised from the background regions. For the MOS spectra, single, double and
triple event patterns were used, while in the PN spectra single and double
events were used. Events with nominal energy $<0.2$ keV were excluded because
the EPIC responses are uncertain at such low energies. We further excluded
events with nominal energy $<0.4$ keV in PN because of the increased noise
levels when double events are selected in this energy range.  Response matrices
for each source were produced using the SAS tasks {\small RMFGEN} and {\small
ARFGEN}. Photons from channels contaminated by the strong instrumental
fluorescent emission lines (Al K in MOS and Cu K in PN) were excluded from both
source and background spectra. 
Spectra from the five observations and three EPIC cameras were combined by
summing the counts from channels with the same nominal energy range. Background
spectra were summed in the same fashion. The response matrices were combined by
averaging the effective areas of the individual response matrices for each
channel and each energy range. The procedure is described in detail in Page,
Davis \& Salvi (2003). The resultant spectra were grouped to a minimum of 20
counts per bin before spectral fitting.

\subsection{Near-infrared/optical observations}

We used the near-infrared camera UFTI on the $3.8$-m United Kingdom InfraRed
Telescope (UKIRT) to make images at $J$ and $K$ (and $H$ for ero5 only;
$H=19.30\pm0.15$ mag) on 2001 February 17 and 2001 March $12-16$. UFTI is a
$1024\times1024$ InSb imager with 0.091 arcsec pixels. Data were taken in
periods of good atmospheric transparency and with seeing of $0.4-1.1$ arcsec.
We used a $3\times3$ grid of dithered positions for each image, allowing the
construction of a sky frame for flatfielding. Integration times are given in
Table~\ref{table:oir}. Calibration was obtained by observing a selection of
UKIRT faint standards (Hawarden et al. 2001) on each night.

The data were analysed with the ORAC-DR data reduction pipeline which automates
the dark subtraction, flat-fielding, resampling and despiking steps. The
resulting mosaics were coadded as required with the STARLINK {\sc CCDPACK}
tasks {\sc ccdalign} and {\sc makemos}. We obtained secure identifications by
calibrating the data astrometrically using bright stars in the
images. Magnitudes, calculated in 2 arcsec diameter apertures, are given in
Table~\ref{table:oir} along with $R-K$ and $J-K$ colours. 
	
\subsection{Millimetre and submillimetre observations}

Submillimetre observations were made at the James Clerk Maxwell Telescope
(JCMT) on the 2001 January $22-24$. We used SCUBA in photometry mode (see
Holland et al. 1999) to make sensitive continuum observations in the 850- and
450-$\mu$m atmospheric passbands. The measured beam sizes at these wavelengths
are approximately $14.5$ and $8.0$ arcsec respectively. The adopted chop throw was
45 arcsec in azimuth. Atmospheric transparency and stability were in the top
quartile of conditions experienced on Mauna Kea; the zenith opacity was
monitored with skydips yielding values of $0.18-0.19$ and $0.8-0.9$ at 850 and
450~$\mu$m respectively. Telescope pointing accuracy was checked on the nearby
blazar $1044+719$ at $25-45$ min intervals. Local pointing offsets were in all
cases better than $2$ arcsec.

Data reduction was performed in the standard manner using the STARLINK package
{\sc surf} (Jenness \& Lightfoot 1998), and flux calibration was made with
respect to the JCMT primary calibrator, Mars. Calibration uncertainties are
estimated to be $\sim5-10$ per cent at 850 $\mu$m and $\sim15-20$ per cent at
450 $\mu$m. The resulting rms noise limits on our coadded photometric
observations give noise equivalent flux densities of $\sim80$ and $\sim800$ 
mJy.Hz$^{-1/2}$ at 850 and 450 $\mu$m respectively, entirely consistent with
both predicted and previously measured values in similar weather
conditions. Submillimetre flux densities are given in Table~\ref{table:radsub}

One source, ero5, was observed at the IRAM 30-m telescope during March 2001 with
the MPifR 37-channel bolometer array operating at 1.25 mm. The integration time
was 3314 secs with the array operating in ON/OFF mode. The observing and data
reduction strategy is analogous to that used with SCUBA at the JCMT. 

\subsection{Radio observations}

The Lockman field was mapped with the National Radio Astronomy
Observatory's (NRAO) Very Large Array (VLA).

At 1.4\,GHz, interference and the presence of bright sources in the primary
beam meant that we had to use the spectral-line correlator
mode, recording data every 5\,s in 3.25 MHz channels, 28 in total, including
left-circular and right-circular polarizations. 3C\,84 and 3C\,286 were used
for flux calibration. The phase/amplitude calibrator, 1035+564, was observed
every hour. A total of 75\,h of integration was obtained --- 70\,h in A
configuration; 5\,h in B configuration. The lengthy data reduction procedure is
described in detail by Ivison et al.\ (2002).  The resulting map has a noise
level of $\sim 4-5$\,$\mu$Jy\,beam$^{-1}$ with 1.4 arcsec resolution.

At 4.9\,GHz, data were taken in the standard continuum correlator mode in C
configuration, with a total bandwidth of 100\,MHz, and reduced according to
the standard {\sc aips} recipes. The resulting map has a noise level of
$\sim 13$\,$\mu$Jy\,beam$^{-1}$ with 5 arcsec resolution. 

Radio source positions measured at 1.4 GHz, together with flux densities and
upper limits, are given in Table~\ref{table:radsub}.

\section{Results}

\subsection{X-ray properties}
\label{sec-xray} 

\begin{figure}
\setlength{\unitlength}{1in}
\begin{picture}(3.25,8.5)
\includegraphics{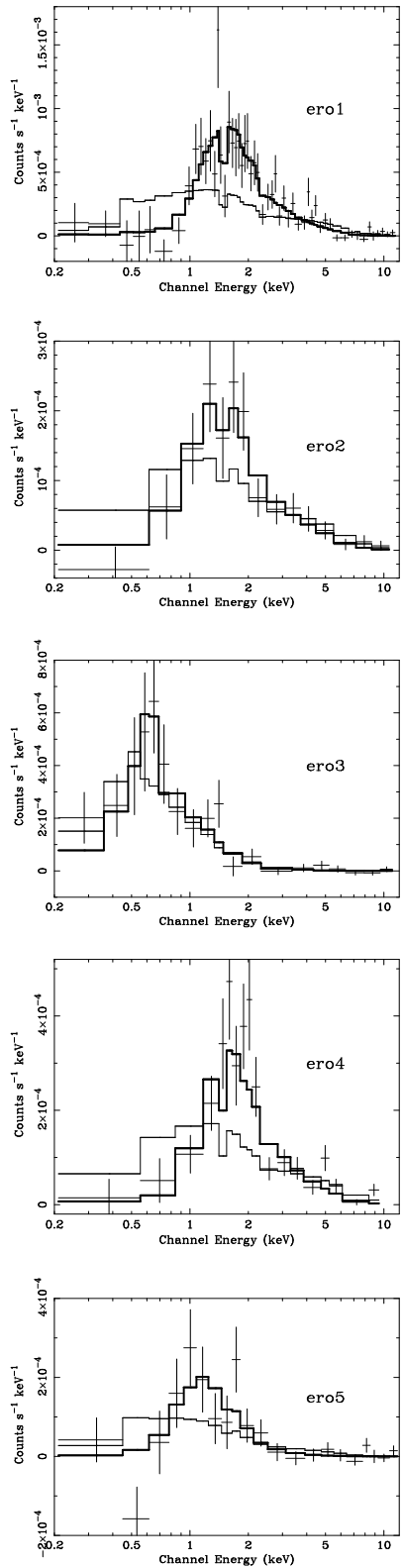}
\end{picture}
\caption[dum]{X-ray spectra of the five EROs. In each case, the thin line shows
the best power law model folded through the response, while the bold line shows
the overall best fit model (a thermal plasma for ero3, an absorbed power law
for the other EROs) folded through the response. The drop in response below 0.4
keV is due to the exclusion of the PN data in this energy range (see Section
3.1).}
\label{fig:xmmspec}
\end{figure}

The X-ray fluxes of the EROs, derived from multi-band source detection and
parameterization are listed in Table~\ref{table:posx}.  It is immediately clear
that ero3 is much softer than the other EROs: it is the only source detected
below 0.5 keV but it is the weakest source at energies higher than 2 keV. As
expected, this difference is borne out by spectral fitting.

\begin{figure*}
\setlength{\unitlength}{1in}
\begin{picture}(6.0,4.25)
\includegraphics{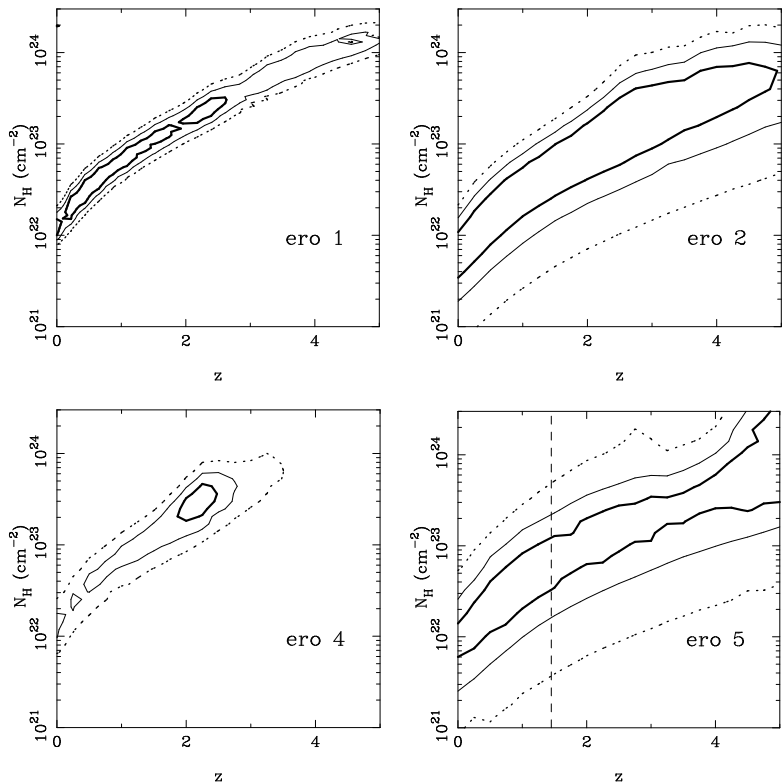}
\end{picture}
\caption[dum]{Confidence intervals for the redshift and absorbing column
density for those EROs with X-ray spectra that are best fitted by an absorbed
power law. The bold solid, thin solid, and dotted contours correspond to 1, 2
and 3\,$\sigma$ respectively. The dashed line on the lower right panel
corresponds to $z=1.45$ for ero5; at 95 per cent confidence level the range of
acceptable column densities is $6.2^{+10.5}_{-4.0}\times10^{22}$.}
\label{fig:nh}
\end{figure*}

\begin{figure}
\setlength{\unitlength}{1in}
\begin{picture}(3.25,2.0)
\includegraphics{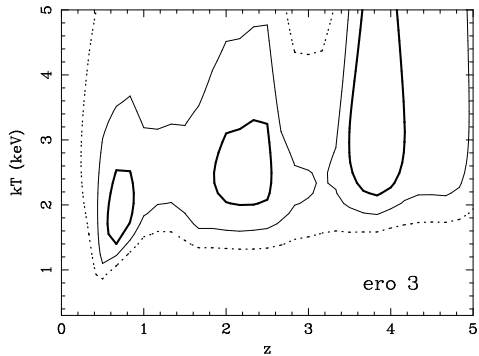}
\end{picture}
\caption[dum]{Confidence interval on temperature and redshift for a thermal
plasma model fit to the X-ray spectrum of ero3. The bold solid, thin solid, and
dotted contours correspond to 1, 2 and 3$\sigma$ respectively.}
\label{fig:thermal}
\end{figure}

Initially, each of the ERO spectra were fitted with a power law model, absorbed
only by the Galactic column in the direction of the Lockman Hole ($6 \times
10^{19}$ cm$^{-2}$; Lockman et al. 1986). The results of these fits (and
those discussed below) are given in Table~\ref{table:xmmfits} and are shown
graphically in Fig.~\ref{fig:xmmspec}. The power law model is an extremely poor
fit to ero1 and ero2, and the fitted power law indices for ero1, ero2 and ero4
are much harder than the spectra of typical unobscured AGN. The power law fits
to ero1, ero2 and ero4 also result in strong residuals which are characteristic
of photoelectric absorption: an excess of counts above 1 keV, and a deficit of
counts below 1 keV. We have therefore re-fitted the spectra using an absorbed
power law model, with the redshift and column density of the absorbing medium
as free parameters; Galactic absorption was included as before.  This results
in much better fits to the spectra of ero1, ero2, ero4 and ero5. In all four
cases the improvement in $\chi^{2}/\nu$ (here $\nu$ is the number of data
points minus the number of free parameters) is significant at $> 99$ per cent
according to the F-test, and the intrinsic power-law spectral indices are now
consistent with those found in normal AGN. The derived column densities shown
in Fig.~\ref{fig:nh} depend heavily on redshift, but probably lie somewhere
between $10^{22}$ and $10^{24}$~cm$^{-2}$. Finally, to see if the X-ray
emission could have a thermal origin, we have attempted to fit the spectra
using a Mekal optically-thin thermal plasma model with Galactic absorption. We
assumed Solar abundances for the plasma but its redshift was a free parameter
in the fits. This model is a poor fit regardless of redshift for ero1, ero2 and
ero4, and for ero5 it is a poorer fit than the absorbed power law model. For
ero3, however, it is a slightly better fit than an absorbed or unabsorbed power
law. The acceptable range of temperatures and redshifts for this source are
shown in Fig.~\ref{fig:thermal}. For each ERO, the best-fit model is
overplotted on the observed spectrum in Fig.~\ref{fig:xmmspec}.

\begin{table*}
\caption{Model fits to the ERO X-ray spectra. The energy index $\alpha$ is
defined here (and throughout the paper) as $S_{\nu} \propto \nu^{\alpha}$. The
errors are quoted at 95\% confidence for one interesting parameter
(i.e. $\Delta \chi^{2}=4$); an `*' indicates that the fit parameter reaches the
limit on its allowed range before $\Delta \chi^{2}=4$. `Prob' is the
probability of obtaining the observed $\chi^{2}/\nu$ or greater if the data
were drawn from the model.}
\label{table:xmmfits}
\begin{tabular}{lccccccccc}
Source & \multicolumn{3}{c}{------------ power law ------------}
& \multicolumn{3}{c}{------absorbed powerlaw ------} 
& \multicolumn{3}{c}{--------- thermal plasma---------}\\
     & $\alpha$ & $\chi^{2}/\nu$ & Prob &
       $\alpha$ & $\chi^{2}/\nu$ & Prob &
        kT (keV) & $\chi^{2}/\nu$ & Prob \\
&&&&&&&&\\
ero1 & $0.3_{-0.1}^{+0.2}$ & $162.3/51$ & $1.6\times 10^{-13}$ &
       $-1.7_{-0.6}^{+0.5}$ & $57.5/49$ & $0.19$ &
       $100^{+*}_{-42}$    & $199.0/51$ & $1.1 \times 10^{-19}$\\
ero2 & $0.3_{-0.3}^{+0.3}$ & $22.1/13$ & $5.4\times 10^{-2}$ &
       $-0.8_{-1.0}^{+0.8}$ & $5.8/11$ & $0.88$ &
       $100^{+*}_{-78}$    & $38.2/12$ & $1.4\times 10^{-4}$\\
ero3 & $-1.5_{-0.6}^{+0.5}$ & $16.4/17$ & $0.50$ &
       $-1.8_{-3.9}^{+0.8}$ & $15.7/15$ & $0.40$ &
       $1.9^{+1.2}_{-0.6}$ & $14.0/16$ & $0.60$ \\
ero4 & $0.3_{-0.2}^{+0.3}$ & $62.9/16$ & $1.7\times 10^{-7}$&
       $-1.5_{-1.2}^{+0.8}$ & $24.3/14$ & $4.2\times 10^{-2}$&
       $100^{+*}_{-40}$    & $90.9/15$ & $6.8\times 10^{-13}$\\
ero5 & $-0.2_{-0.6}^{+0.7}$ & $32.5/19$ & $2.7\times 10^{-2}$&
       $-2.6_{-3.9}^{+1.6}$ & $18.5/17$ & $0.356$ &
       $1.5^{+*}_{-0.6}$   & $29.6/18$ & $4.1\times 10^{-2}$\\
\end{tabular}
\end{table*}

\subsection{Radio to near-infrared properties}

\label{sec-rnir}

In this section we use our multicolour photometric and imaging data to
investigate the nature of our five X-ray selected EROs.

\begin{figure}
\setlength{\unitlength}{1in}
\begin{picture}(3.0,3.25)
\includegraphics{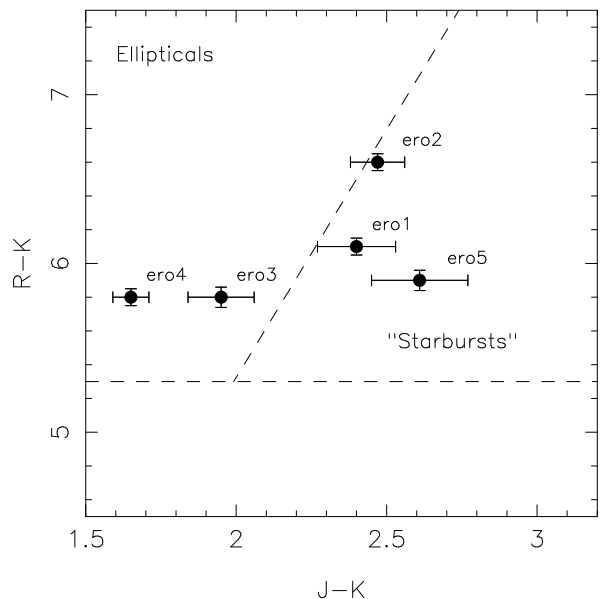}
\end{picture}
\caption[dum]{Colour-colour diagram. The diagonal dashed line separates
passively evolving elliptical galaxies from dusty `starburst' galaxies
(Pozzetti \& Mannucci 2000). Our EROs are chosen to lie above the horizontal
dashed line at $R-K>5.3$.}
\label{fig:pm}
\end{figure}

Since an old stellar population will show a break in the 4000\,\AA\, region
whereas a dusty starburst (or dusty AGN) will have a smoother and shallower
overall spectrum it has been proposed that the two populations will fall in
different regions of, for example, a $R-K$ vs $J-K$ diagram (Pozzetti \&
Mannucci 2000). The first indications are that this is indeed the case, with
known elliptical and starbursting EROs falling in the regions predicted by
these spectral synthesis models (Pozzetti \& Mannucci 2000). In addition, it
has been shown that EROs detected in a very deep radio map (where the radio
flux is assumed to be connected with star formation -- see below) have redder
$J-K$ colours on average than those that were not detected (Smail et
al. 2002). In the same study, spectral synthesis models fitted to broad-band
photometric data showed that those sources best-fitted by dusty starburst
models tended to have redder $J-K$ colours than those fitted by an evolved
stellar population (Smail et al. 2002).  In Fig.~\ref{fig:pm} we use our
optical and near-infrared photometry to produce a $R-K$ vs $J-K$ colour-colour
diagram. Three of the five sources have clear-cut classifications according to
this diagnostic tool; ero5 is a dusty `starburst' source, while ero3 and ero4
are `ellipticals'. In addition, ero1 may also be dusty but ero2 falls on the
boundary between the two classifications.
 
\begin{figure*}
\setlength{\unitlength}{1in}
\begin{picture}(3.0,2.0)
\includegraphics{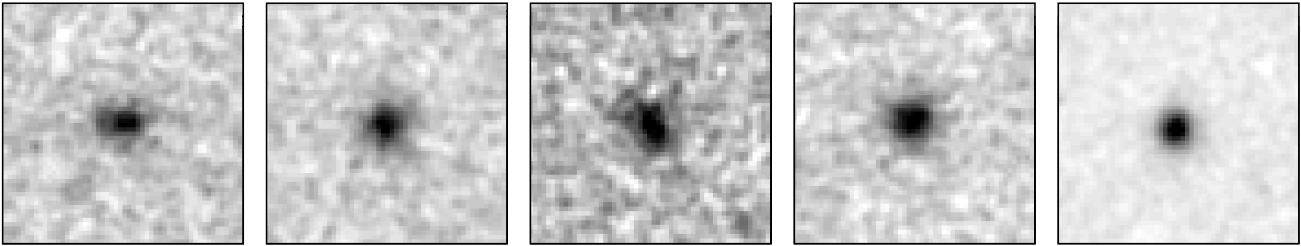}
\end{picture}
\caption[dum]{$K$-band images smoothed with a Gaussian of 2 pixels FWHM. From
left to right are ero1, ero2, ero3, ero4 and ero5. Each panel is $5 \times 5$
arcsec.}
\label{fig:irimages}
\end{figure*}

\begin{figure}
\setlength{\unitlength}{1in}
\begin{picture}(3.0,3.25)
\includegraphics{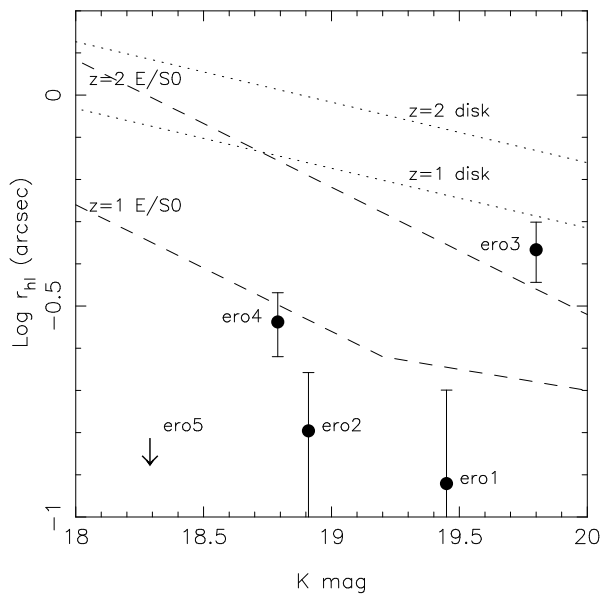}
\end{picture}
\caption[dum]{Half-light radii against $K$ magnitude. The dotted and dashed lines
are models taken from Roche et al. (2002).}
\label{fig:hlk}
\end{figure}

Small $K$-band images centred on the EROs are presented in
Fig.~\ref{fig:irimages}.  The images are not deep enough to allow profile
fitting but we were able to extract half-light radii ($r_{\rm hl}$) by
comparing the EROs with stars in their respective images, i.e. we determined
the full width half maximum of the ERO and the best star in the frame,
deconvolved them assuming Gaussian sources and then translated the results into
half-light radii. The results are presented in Table~\ref{table:oir}, and we
plot $r_{\rm hl}$ versus $K$ magnitude in Fig.~\ref{fig:hlk} together with
models taken from Roche et al. (2002). These models are based on the
size-luminosity relations of local spheroidal (E/S0) and spiral (disk) galaxies
combined with passive luminosity evolution. The results for ero3 and ero4 agree
with those found above from their photometric colours; they are consistent with
passively evolving elliptical or spiral galaxies, ero4 with a $z=1$ E/S0 and
ero3 with either a $z=2$ E/S0 or a $z=1$ disk. Inspection of the images shown
in Fig.~\ref{fig:irimages} shows that ero3 might have an elongated morphology
typical of an inclined spiral galaxy, although its size and morphology might
alternatively be ascribed to an ongoing merger.  The other three EROs are
inconsistent with all of the models - this is particularly the case for ero5
which is indistinguishable from the stars on its image. As a comparison, {\em
Hubble Space Telescope\/} imaging data (Smith et al. 2002) show that only 3 of
62 $K$-band selected EROs with $R-K>5.3$ (1/27 with $R-K>6.0$) have sizes
smaller than 0.16 arcsec (that of ero2).  Similarly, ero1, ero2 and ero5 are
smaller than practically all of the 32 $K\leq19.5$ EROs selected from $R$- and
$K$-band imaging data by Roche et al. (2002).


If any of the X-ray selected EROs are luminous, heavily dust reddened
starbursts they will show signatures of star formation in other wavebands.  In
particular, we expect them to have strong emission in the submillimetre band
from warm dust heated by young stars, and synchrotron radiation in the radio
band resulting from supernovae.  If, instead, the sources are red because they
have a redshifted, evolved stellar population, they will not show these
characteristics. This simple picture, however, is complicated by the presence
of an AGN which can produce synchrotron radiation at radio wavelengths, and at
submillimetre wavelengths can heat circumnuclear dust.  Of the five sources
only one, ero5, is detected at 850~$\mu$m with SCUBA
(Table~\ref{table:radsub}). Three of them (ero1, ero4 and ero5) are detected at
one or both radio frequencies (Table~\ref{table:radsub}). All three of the
radio detected sources are unresolved at the 1.4 arcsec resolution of the
1.4-GHz VLA map (Table~\ref{table:radsub}).

We discuss ero5 in detail in Appendix~\ref{sec:ero5} where we conclude that it
is probably a radio-quiet AGN and that the submillimetre emission is from warm
dust. We can investigate the heating source of the dust by comparing the
bolometric luminosity emitted by the AGN ($L_{AGN}$) to that emitted in the
far-infrared ($L_{FIR}$). Assuming an intrinsic X-ray spectral index of $-1$
and using the absorbed power law fit reported in Section~\ref{sec-xray}, the
flux emitted in the $0.5-2$~keV band, corrected for absorption is
$3.0^{+1.7}_{-1.3}\times10^{-15}$~erg\,s$^{-1}$. If 3 per cent of the AGN
luminosity is emitted in the $0.5-2$~keV band (Elvis et al. 1994) then
log$(L_{AGN})=11.55^{+0.19}_{-0.25}\ L_{\odot}$ at $z=1.45$. We assume further
that ero5 has the same far-infrared spectrum as Mrk~231, an X-ray absorbed AGN
in the local Universe with log$(L_{FIR})=12.59~L_{\odot}$ (see Page et al. 2001
and references therein). If observed at $z=1.45$ Mrk~231 would have an
850~$\mu$m flux density of 1.9~mJy.  Scaling with ero5 we derive
log($L_{FIR})=12.95\pm0.10~L_{\odot}$, large enough to classify ero5 is an
ultraluminous infrared galaxy or ULIRG. The ratio
$L_{FIR}/L_{AGN}=25^{+13}_{-15}$ suggests that the far-infrared luminosity of
ero5 is powered by star formation, in which case the star-formation rate (SFR)
can be calculated from SFR(M$_{\rm\odot}$yr$^{-1})=L_{FIR}/5.8\times10^9$
(Kennicutt 1998) giving SFR$\sim1500$~M$_{\rm\odot}$yr$^{-1}$.

The other two sources detected at radio wavelengths are fainter than ero5 by
factors of $\sim20$ (ero1) and $\sim45$ (ero4) at 1.4~GHz, and are thus also
unlikely to be radio-loud given their similar optical/near-infrared
magnitudes. Their radio emission could, however, be associated with strong
star formation and/or weak AGN activity.
Taking advantage of the opposite (and strong in the submillimetre case)
K-corrections in the two bands, a number of studies have used the radio to
submillimetre spectral index as a redshift indicator (e.g., Carilli \& Yun
1999; Dunne, Clements \& Eales 2000). Assuming a $2\sigma$ upper limit of 2.8
mJy, ero1 has a spectral index $\alpha_{1.4}^{350}<0.58$ leading to rough
redshift constraints $z<1.5$ and $z<1.0$ respectively from the two studies
cited above. Similarly, ero4 has $\alpha_{1.4}^{350}<0.83$ leading to $z<3.5$
and $z<2.2$ respectively, and consistent with that deduced from
Fig.~\ref{fig:hlk}. However, if the AGN contributes to the radio emission then
higher redshifts (up to $z=6$ or so) are possible. We can put a rough limit on
their SFRs by assuming that all of the radio emission has a starburst origin
and that they are at $z=1$. Then, SFR(M$_{\rm
\odot}$yr$^{-1})=5.9\pm1.8\times10^{-22}L_{\rm 1.4 GHz}$ where $L_{\rm 1.4
GHz}$ is the 1.4-GHz luminosity in units of W\,Hz$^{-1}$ (Yun, Reddy \& Condon
2001). The calculated SFRs are $\leq340$ and $\leq90$ M$_{\rm \odot}$yr$^{-1}$
for ero1 and ero4 respectively.
 
Two sources, ero2 and ero3, are undetected in both the radio and submillimetre
bands. Their radio detection limits give SFRs of $<80$ and $<50$
M$_{\rm \odot}$yr$^{-1}$ respectively at $z=1$.

\section{Discussion}

\label{sec-disc}

\begin{table*}
\footnotesize
\centering
\def\baselinestretch{1}                                     
\caption[dum]{\small Summary of source properties and proposed classification.}
\label{table:summary}
\vspace*{0.1in}
\begin{tabular}{cccccl}
Name & Best fit& colour & $K$-band size & submm/radio SFR & conclusion \\
     & X-ray spectrum & diagnostic & diagnostic & (M$_{\odot}$\,yr$^{-1}$)$^a$&\\
     & & & & & \\
ero1 & absorbed power law & starburst & AGN & $340$ & absorbed AGN with starburst.\\
ero2 & absorbed power law & ambiguous & AGN & $<80$ & absorbed
     AGN. Quiescent host with circumnuclear torus?\\
ero3 & thermal plasma & elliptical & elliptical & $<50$ & giant elliptical galaxy \\
ero4 & absorbed power law & elliptical & elliptical & $90$ & absorbed
     AGN. Quiescent host with circumnuclear torus? \\
ero5 & absorbed power law & starburst & AGN & $1500$ & absorbed
     AGN with starburst. ULIRG. \\ \\
\multicolumn{6}{l}{\small $^a$SFRs are calculated assuming $z=1$ except for
     ero5 which has a measured redshift of $z=1.45$}
\end {tabular}
\end{table*}

The results from the previous section and the proposed source
classifications discussed below are summarized in Table~\ref{table:summary}.

Our X-ray spectral analysis shows that ero1, ero2, ero4 and ero5 almost
certainly contain absorbed AGN. The optical/near-infrared colours suggest that
absorbed AGN occur in both types of ERO: dust-reddened starbursts (e.g. ero5),
and EROs that are dominated by an old stellar population (e.g. ero4). The
$K$-band half-light radius of ero4, as well as the optical/near-infrared colours,
are consistent with an early type galaxy at $z\sim 1$ assuming passive
luminosity evolution; this suggests that the AGN in this source contributes
very little to its optical/near-infrared light.

In contrast to ero4, the two `dusty starburst' EROs and the source with an
ambiguous colour classification (ero2) are unusually compact in the $K$-band
images, suggesting that their near-infrared emission includes a significant
contribution from the absorbed AGN. The column densities measured in the X-ray
($< 10^{23} {\rm cm^{-2}}$ for $z\sim 1$) do not preclude such a possibility.
Indeed, the dust-reddened optical/near-infrared colours of AGN are
indistinguishable from the colours of young, massive stellar populations, and
therefore it is possible that these three sources were selected as EROs because
they contain dust obscured AGN rather than dusty star formation.
This potential AGN contribution means that the optical/near-infrared colours
alone are not sufficient to determine the nature of the host galaxies in these
three objects.  However, the submillimetre detection of ero5 suggests that this
source {\em is\/} a dusty starburst galaxy. In Section~\ref{sec-rnir} we argue
that it is a ULIRG forming stars at a rate of
$\sim1000-1500$~M$_{\rm\odot}$yr$^{-1}$. The detection of radio emission from ero1
suggests that this source is also a dusty, star-forming galaxy. However, the
lack of any detectable radio or submillimetre emission from ero2 suggests that
like ero4, it could be an absorbed AGN hosted by a quiescent, early type
galaxy.


It is thus apparent that absorbed AGN are found in both types of ERO:
elliptical galaxies at $z\sim 1$ and high-redshift dusty starbursts.  EROs such
as ero2 and ero4, which correspond to the first type, have properties that are
consistent with a normal galaxy containing an AGN which is viewed through an
obscuring torus, as expected from AGN unified schemes (Antonucci 1993), and
hypothesized in large numbers to explain the X-ray background (e.g. Comastri et
al. 1995). In these cases, the dust is located close to the AGN itself, and may
be hot. At least two hard X-ray sources discovered by {\em Chandra\/} have been
shown to exhibit these properties (Wilman, Fabian \& Gandhi 2000; see also
Deane \& Trentham 2001).  EROs such as ero1 and ero5, which correspond to the
second type, appear to be undergoing vigorous star formation. In these sources
some obscuration of the AGN could be due to warm dust and gas which is
associated with the star-forming regions, rather than an obscuring torus. Such
sources may be AGN at an earlier evolutionary phase than the unobscured AGN
population (e.g. Fabian 1999).  Whatever the host galaxy, if the AGN makes a
significant contribution to the $K$-band flux the position of the source on the
colour-colour diagram could be shifted relative to position of the host galaxy,
and we caution against applying photometric redshift techniques to Compton-thin
AGN (e.g. Crawford et al. 2002).

If we assume that the $5-12$~keV flux density is little affected by absorption
then the AGN are characteristic of those found at the break of the X-ray
luminosity function ($L\star$) to within a factor of $\sim5$. For example, ero5
at $z=1.45$ has an intrinsic luminosity, $L_{X}\sim10^{44.1}$ erg~s$^{-1}$,
while ero1 at $z=3$ would have $L_{X}\sim10^{45.2}$ erg\,s$^{-1}$ compared to
$L\star$ $(z=2) \sim10^{44.5}$ erg\,s$^{-1}$ (Page et al. 1997). This result
implies that four out of five of our EROs contain absorbed AGN with equivalent
luminosities to those that make up the bulk of the unobscured component of the
extragalactic X-ray background.

The only ERO in our sample that does not appear to have a significant
contribution from an AGN is ero3. If it does contain an AGN, then the X-ray
spectral fitting suggests that it cannot be absorbed by a significant amount of
material. Its diffuse, extended near-infrared morphology, however, shows that
such an unabsorbed AGN must be weak, because it makes a negligible contribution
to the $K$-band light. Alternative sources of the X-ray emission are powerful
starburst activity or halo emission from a massive elliptical galaxy in a deep
gravitational potential. The former case is unlikely for a number of
reasons. Firstly, ero3 was undetected in both the submillimetre and radio bands
showing that star formation is not occurring on a massive scale; the limit
derived from the deep radio map is SFR $<50$ M$_{\rm \odot}$yr$^{-1}$ at $z=1$.
Secondly, its X-ray flux corresponds to a luminosity $L_{X}\sim10^{43.5}$
erg~s$^{-1}$ ($z=1$) or $L_{X}\sim10^{44.1}$ erg~s$^{-1}$ ($z=2$). This
luminosity is too large for almost any conceivable starburst; the brightest
X-ray starburst known is NGC~3256 with $L_{X}=10^{42.4}$ erg~s$^{-1}$ (Moran,
Lehnert \& Helfand 1999), more than an order of magnitude less luminous than
ero3. 
The temperature that we obtain from the fit to the X--ray spectrum ($\sim$
1--3 keV, see Fig. 3) and the luminosity are, however, within the range of
values found for cD galaxies locally (O'Sullivan, Forbed \& Ponman 2001,
Matsushita 2001). This is consistent with the optical/IR colour and size 
results discussed above and suggests that ERO3 is likely to be the dominant
galaxy of a cluster.

These results are in good agreement with those deduced from the {\em Chandra\/}
Deep Field North Survey (Alexander et al. 2002). They claim that X-ray selected
EROs that were detected in the hard band are absorbed AGN (9 out of 13
sources), while the 4 sources detected only in the soft band have X-ray
emission from less energetic processes.  In particular, a stacking analysis of
those EROs with $I<24.1$ that were not individually detected resulted in a
statistical detection in the soft band. They argue that this result is
consistent with that expected if the majority of these sources are typical
elliptical galaxies.  X-ray spectral fitting shows convincingly that 4/5 of our
sources are absorbed AGN, and that the other source is likely to be an
elliptical galaxy, albeit a very luminous example.

Although our sample is small, two of the absorbed AGN are likely to contain
substantial quantities of warm dust. We speculate that the other absorbed AGN
might contain only hot circumnuclear dust. These properties are in keeping with
the predictions of models for the X-ray background (Fabian \& Iwasawa
1999). This result might also suggest that EROs detected as counterparts to
faint submillimetre galaxies (e.g. Smail et al. 1999) have a reasonable chance
of being associated with AGN, although we have already noted that only $\sim4$ per
cent of EROs in the Lockman Hole sample are X-ray sources (Section~\ref{sec-sample}).
Similarly, Alexander et al. (2002) find that only
$14^{+11}_{-7}$ per cent of the $K\leq20.1$ ERO population are obscured AGN to
the depth of their survey. Another study (Smail et al. 2002) suggests that this
number could be as low as 2 per cent although their X-ray data are not as deep
as those presented by Alexander et al. (2002).  A recent comparison between
faint \xmm sources and submillimetre galaxies selected from the bright end of
the number counts showed an overlap of 4 out of 16, with only one of these
sources identified as both an X-ray source and an ERO (Ivison et
al. 2002). However, these limits only constrain the fraction of EROs which
contain Compton-thin AGN. It will require considerably deeper X-ray
observations to detect a population of Compton-thick EROs with X-ray
luminosities similar to, or lower than, the 5 EROs studied in this paper.

\section{Conclusions}

\begin{enumerate}

\item We present multifrequency follow-up observations and analysis for 5 out
of 6 EROs identified in the 100 ks \xmm observation of the Lockman Hole. They
have enough counts in the X-ray band to allow spectral fitting. Four of them
are best-fitted by an absorbed power law model with redshift dependent column
densities of $10^{22}-10^{24}$~cm$^{-2}$. They are most probably Compton thin,
obscured AGN. The fifth source has a best fit thermal spectrum and might be a
massive elliptical galaxy with X-ray emission associated with its halo.

\item The EROs have optical/near-infrared colours consistent with dusty sources
or normal elliptical galaxies. $K$-band imaging data show that those sources
with elliptical colours have sizes predicted for elliptical galaxies at
$1<z<2$. The other three sources are smaller than the model predictions; it is
likely that in these cases the host galaxy is dominated by light from the AGN,
consistent with them being Compton-thin AGN.

\item The dust content of the EROs is investigated with sensitive observations
in the submillimetre and radio bands. Of the two AGN with dusty colours, one of
them is detected at 850~$\mu$m which we interpret as emission from warm dust
heated by young stars. The other is detected at 1.4 and 4.9~GHz suggesting that
it also is a galaxy undergoing a burst of star formation, albeit at a less
spectacular rate. Of the two sources with elliptical-like colours, one of them
(that with the best-fit thermal X-ray spectrum) is undetected in the
submillimetre and radio bands. The other is a faint radio source -- this radio
emission might be attributable to the AGN. We argue that the fifth source, an
AGN with an ambiguous colour classification, has relatively little warm dust,
and that the nucleus makes a significant contribution to the $K$-band light.

\item We suggest that the AGN with non-dusty colours might contain hot dust
that provides viewing angle dependent obscuration (e.g. a circumnuclear torus),
while the dusty AGN might be obscured by warm dust and gas associated with starburst
activity.  The latter might be in an earlier evolutionary phase than the AGN
hosted by quiescent, early type galaxies.

\item The EROs that host AGN have luminosities within a factor of 5 of
those found at the break of the unabsorbed X-ray luminosity function, i.e. they
have luminosities equivalent to AGN that make up the bulk of the
unobscured X-ray background.

\end{enumerate}  

\section*{ACKNOWLEDGMENTS}

The James Clerk Maxwell Telescope is operated by the Joint Astronomy Centre in
Hilo, Hawaii on behalf of the parent organizations PPARC in the United Kingdom,
the National Research Council of Canada and The Netherlands Organization for
Scientific Research. The United Kingdom Infrared Telescope is operated by the
Joint Astronomy Centre on behalf of PPARC.  Based on observations obtained with
{\em XMM-Newton}, an ESA science mission with instruments and contributions
directly funded by ESA member states and the USA (NASA).  NRAO is operated by
Associated Universities Inc., under a cooperative agreement with the National
Science Foundation. J.A.S. acknowledges support from PPARC. I.R.S. acknowledges
support from the Royal Society and the Leverhulme Trust.

\bsp

\appendix\section{What is ero5?}
\label{sec:ero5}

\begin{figure}
\setlength{\unitlength}{1in}
\begin{picture}(3.0,4.6)
\includegraphics{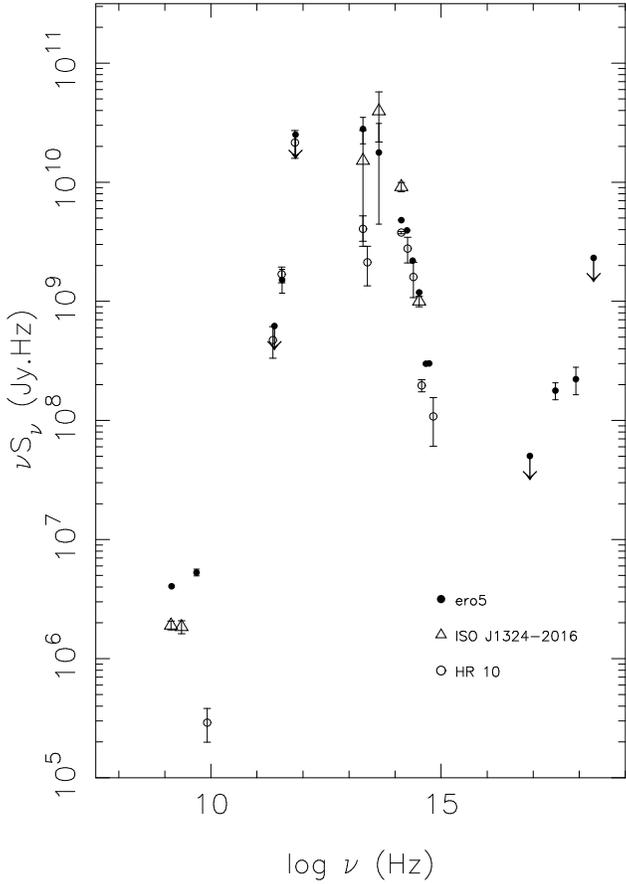}
\end{picture}
\caption[dum]{Spectral energy distribution of ero5 plotted for comparison
against those of HR~10 (Dey et al. 1999; Elbaz et al. 2002 [mid-infrared]) and
ISO J1324$-$2016 (Pierre et al. 2001). No shifts in absolute flux level have
been applied to the data.}
\label{fig:ero5}
\end{figure}

The spectral energy distribution of ero5 is plotted in Fig.~\ref{fig:ero5}
where we also show those of the ULIRG ERO J164502+4626.4 otherwise known as
HR~10 (Dey et al. 1999) and the dusty quasar ISO J1324$-$2016 (Pierre et
al. 2001), both of which are EROs.  The comparison sources, both at $z \sim
1.5$, are plotted in the observed frame, and have not been shifted in flux
level. The mid-infrared data are from {\em ISOCAM\/} observations of the
Lockman Hole which show that ero5 is one of the brightest $15\mu$m sources in
the X-ray detected sub-sample (Fadda et al. 2002).  There is a striking
similarity between ero5 and HR~10 in the millimetre and near-infrared optical
regime but ero5 is at least an order of magnitude brighter at radio wavelengths
where it is more similar to ISO J1324$-$2016.  The {\em ISOCAM\/} data have
large uncertainties but nevertheless there is reasonably good evidence that the
AGN have rising spectra between the $K$-band and the mid-infrared whereas the
starburst-like HR~10 shows the opposite behaviour.

The millimetre--submillimetre spectral index of $>2.4$ practically rules out
a self-absorbed synchrotron origin for the submillimetre radiation
unless the source varied in between the observations which we consider to be
unlikely.  Although the spectrum is just consistent with that expected for a
homogeneous synchrotron source in the optically thick regime (maximum slope of
2.5) such sources have never been observed with turnovers in the far-infrared
(see e.g. discussion in Hughes et al. 1993). The most likely interpretation is
thus that the submillimetre flux is due to thermal emission from dust.

The radio emission from ero5 is likely to include a large contribution from the
AGN.  The spectral index between the radio and submillimetre bands is
$\alpha_{1.4}^{350}=0.07$ which would place the source at $z<0.3$ if the radio
emission is attributed to star formation alone (e.g. Carilli \& Yun 1999;
Dunne, Clements \& Eales 2000). Turning this argument around, and assuming that
all of the submillimetre emission is from dusty star formation, the predicted
1.4~GHz flux density in the observer's frame is approximately $30-300~\mu$Jy at
$z=1.45$. This result implies that most of the $\sim3$~mJy emitted at 1.4~GHz
is from processes other than star formation. 

Could the source be radio-loud?  Our 1.4-GHz radio image constrains the source
size to be $<0.5$ arcsec.  At $z=1.45$ the projected size must thus be smaller
than about 8.6 kpc (or about 12.1 kpc if $q_0=0$) if it is to remain unresolved
in our radio images.  The data presented by Blundell, Rawlings \& Willott
(1999) show that classical double radio sources have linear sizes ranging from
several thousands of kpc to less than one kpc so our size constraints are
consistent with a radio galaxy or mis-aligned quasar. Another diagnostic for
radio-loudness is the two-point spectral index ($\alpha_{ro}$) measured in the
rest frame between 4.86~GHz and the $R$-band, where radio-loud sources have
indices $<-0.35$ (Zamorani et al. 1981). We can K-correct the radio data using
the observed spectral index between 4.86 and 1.4~GHz of $-0.81\pm0.06$, and the
observed $K$-band flux corresponds approximately to the rest frame $R$-band
flux. The calculated index is $\alpha_{ro}=-0.39$ which falls just into the
radio-loud regime. However, only 0.5 mags of extinction would be required at $K$
to place the source on the other side of the divide. Given that ero5 is 
extremely dusty we consider this very likely to be the case, and so it is most
probably radio-quiet according to this diagnostic. We note that the
observed radio flux density and radio spectral index are consistent with
those found for radio quiet quasars (Barvainis, Lonsdale \& Antonucci
1996). The emission is probably optically thin synchrotron radiation.

\end{document}